\documentclass[12pt,preprint]{aastex} 
\begin{document} 
\title{Infrared Spectroscopy of Symbiotic Stars. IV. \\ V2116~Ophiuchi/GX~1+4, 
The Neutron Star Symbiotic
} 

\author{KENNETH H. HINKLE} 
\affil{National Optical Astronomy Observatory\altaffilmark{1},\\ 
P.O. Box 26732, Tucson, AZ 85726-6732} 
\email{hinkle@noao.edu} 
 
\author{FRANCIS C. FEKEL} 
\affil{Tennessee State University, Center of Excellence in Information 
Systems, \\
3500 John A. Merritt Boulevard, Box 9501, Nashville, TN 37209}  
\email{fekel@evans.tsuniv.edu} 

\author{RICHARD R. JOYCE} 
\affil{National Optical Astronomy Observatory,\\ 
P.O. Box 26732, Tucson, AZ 85726-6732} 
\email{joyce@noao.edu} 
  
\author{PETER R. WOOD} 
\affil{Research School of Astronomy and Astrophysics, Mt. Stromlo Observatory,\\
Australian National University, Cotter Road, Weston Creek, ACT 2611,  
Australia} 
\email{wood@mso.anu.edu.au} 
 
\author{VERNE V. SMITH} 
\affil{National Optical Astronomy Observatory,\\
P.O. Box 26732, Tucson, AZ 85726-6732}
\email{vsmith@noao.edu} 
 
\author{THOMAS LEBZELTER} 
\affil{Institut f\"ur Astronomie, T\"urkenschanzstrasse 17, A-1180 Wien,  
Austria} 
\email{lebzelter@astro.univie.ac.at} 
  
\altaffiltext{1}{Operated by Association of Universities for Research in  
Astronomy, Inc., under cooperative agreement with the National Science  
Foundation} 
 
\begin{abstract} 
We have computed, based on 17 infrared radial velocities, the first set
of orbital elements for the M giant in the symbiotic binary V2116
Ophiuchi.  The giant's companion is a neutron star, the bright X-ray
source GX~1+4.  We rule out the previously proposed period of 304 days,
and instead, find an orbital period of 1161 days, by far the longest of
any known X-ray binary.  The orbit has a modest eccentricity of 0.10
with an orbital circularization time of $\lesssim5 \times 10^6$ years.
The large mass function of the orbit significantly restricts the mass
of the M giant.  Adopting a neutron-star mass of 1.35 M$_{\sun}$, the
{\it maximum} mass of the M giant is 1.22 M$_{\sun}$, making it the
less massive star.  Spectrum synthesis analysis of several infrared
spectral regions results in slightly subsolar abundances for most
metals.  Carbon and nitrogen are in the expected ratio resulting from
the red-giant first dredge-up phase.  The lack of $^{17}$O suggests
that the M-giant has a mass less than 1.3 M$_{\sun}$, consistent with
our maximum mass.  The surface gravity and maximum mass of the M giant
result in a radius of 103 R$_{\sun}$, much smaller than its estimated
Roche lobe radius.  Thus, the mass loss of the red giant is via a
stellar wind.  These properties argue that the M giant is near the tip
of the first-ascent giant-branch.  Although the M giant companion to
the neutron star has a mass similar to the late-type star in low-mass
X-ray binaries, its near-solar abundances and apparent runaway velocity
are not fully consistent with the properties of this class of stars.
Thus, in many ways this symbiotic and X-ray binary system is unique,
and various scenarios for its possible evolution are discussed.
\end{abstract}


\section{INTRODUCTION}

The star \object[V2116 Oph]{V2116 Ophiuchi} = GX 1+4 [$\alpha$ =
17$^h$~32$^m$~02\fs16 $\delta$ = $-$24\arcdeg~44\arcmin~44\farcs0
(2000), $V$ = 18.4 mag] is unique among currently known symbiotic stars
because the companion to the late-type giant is a neutron star rather
than a white dwarf or main sequence star.  \citet{letal71} identified
the object as a very bright, hard X-ray source, GX~1+4.  \citet{d76}
found GX~1+4 to be an X-ray variable with a period of two minutes,
suggesting that it is a slow pulsar.  Early observations of its period
changes led to the suggestion that the variations resulted from Doppler
shifts associated with orbital motion \citep{betal76}.  However, it was
later determined that the spin up and spin down of the neutron star
resulted from variable mass accretion \citep{detal81, retal82,
metal88}.  While the present paper is only indirectly concerned with
the neutron star at the heart of GX~1+4, we note that it is an
extraordinary object.  The rate of change of its rotation period is the
greatest of any known pulsar, and the magnetic field, perhaps as large
as B$_\circ$\,$\sim$\,3$\times$10$^{13}$G, is among the largest
measured in any astronomical object \citep{c97}.

Confirmation of the GX~1+4 optical counterpart with an 18th visual
magnitude star came from both a precise ROSAT position \citep{pfs95}
and the detection of optical pulsing and flickering from the candidate
star \citep{jetal97}.  \citet{gf73} had previously shown that the
suspected optical counterpart is an infrared source.  The system is
situated in the direction of the galactic center (l = 1.94\arcdeg, b =
4.79\arcdeg), and is significantly reddened, with $A_v$\,$\sim$\,5 mag
\citep{detal77, setal96, cr97}.  Distance estimates, depending on the
evolutionary state of the red giant, range from three to 15 kpc
\citep{cr97}.  Based on optical spectra, \citet{detal77} suggested that
GX~1+4 = V2116~Oph\footnote{ In this paper we will refer to the
symbiotic system as V2116~Oph and reserve the name GX~1+4 for the X-ray
source.  However, V2116~Oph is classified by the GCVS as a variable
X-ray source and the names are fully synonymous.} \citep{ketal81} is a
symbiotic star.  They also noted that the emission-line strengths
indicate the presence of a circumstellar envelope of radius $\sim$860
R$_\odot$ ($\sim$6$\times$10$^{8}$ km).  In addition, \citet{detal77}
concluded that if the orbital separation is the size of the
circumstellar envelope, the orbital period would be years.

\citet{setal96} observed the optical spectrum at a resolving power of
$\sim$800 and found strong emission lines of hydrogen, neutral helium,
and neutral oxygen as well as O~III, Fe~VII and Fe~X forbidden
emission.  Absorption bands of TiO and VO are conspicuous in the red,
leading to an M5~III spectral type classification for the late-type
star in the system \citep{setal96, cr97}.  Because of the large amount
of reddening and the energy distribution of the M giant, V2116~Oph is
much brighter in the infrared, $K$\,$\sim$\,8~mag, than at optical
wavelengths.  \citet{betal97} and \citet{cvkl98} observed the
near-infrared spectrum of V2116~Oph at resolving powers of up to 4000.
In contrast to the complex optical spectrum, except for strong emission
lines from H~I Brackett $\gamma$ and He~I 10830 \AA\, the infrared
spectrum is typical of a cool M giant.  \citet{cvkl98} interpreted the
He~I line as an outflow of $\sim$300 km~s$^{-1}$ from the M giant.
However, \citet{ketal99} noted similar line widths for optical-region
emission lines and argued that the broad emission lines arise in gas
that is gravitationally bound to the neutron star.

Binary systems containing a red giant are required to have a long
orbital period.  \citet{detal81} noted that a period of at least 120
days is necessary for {\it a}~sin~{\it i} to exceed the stellar radius,
which places V2116~Oph in a unique X-ray binary niche.  There have been
several attempts to determine the orbital period by making the
assumption that mass transfer from the M giant to the neutron star is
enhanced at periastron.  This mass transfer will alter the rotation
period of the neutron star, and hence the pulse rate, so a statistical
analysis of the X-ray pulses should show peaks at the orbital period.
From such analyses \citet{cetal86} and \citet{petal99} concluded that
the orbital period is $\sim$304 days.

In this paper we exploit the typical M-giant near-infrared spectrum
first described by \citet{betal97}.  Our high-resolution, infrared
spectra of V2116~Oph span nearly five years, enabling us to determine
the first spectroscopic orbit of the red giant.  We also present an
evaluation of the surface abundances of the M giant.  The neutron star
in the V2116~Oph binary system is the remnant of a supernova.  The
survival of the binary is remarkable given that the supernova must have
been catastrophic for both the star we now see as the M giant as well
as the surviving remnant.  Our results enable us to further investigate
the properties of the system and improve our understanding of the red
giant member as well as the evolution of this neutron star binary.

\section {OBSERVATIONS AND REDUCTIONS}

Our 17 spectroscopic observations of V2116~Oph were obtained with four
telescopes at three different observatories (Table~1).  From 1999 July
through 2001 March we obtained spectra at the Kitt Peak National
Observatory (KPNO) with the Phoenix cryogenic echelle spectrograph at
the f/15 focus of both the 2.1\,m and the 4\,m Mayall telescopes.   A
complete description of the spectrograph can be found in
\citet{hetal98}.  The widest slit was used, which results in a
resolving power of $\sim$50,000.  The KPNO observations were centered at
several wavelengths between 1.55~$\mu$m and 1.66~$\mu$m.

From 2001 May through 2002 August spectra were obtained in the southern
hemisphere with the 1.88\,m telescope and coud\'e spectrograph system at
the Mt. Stromlo Observatory (MSO).  The detector was an infrared
camera, NICMASS, developed at the University of Massachusetts.  We
obtained a 2-pixel resolving power of 44,000 at a wavelength of
1.63~$\mu$m.  A more complete description of the experimental setup may
be found in \citet{jetal98} as well as in \citet{fetal00}.  The
detector and electronics were previously used for our survey of
northern symbiotics, carried out with the coud\'e feed telescope at
KPNO.

The devastating bush fire of 2003 January destroyed both the 1.88\,m
telescope at MSO and our infrared NICMASS camera.  As a result, from
2003 April through 2004 April we obtained additional observations with
the 8\,m Gemini South telescope, El Pachon, Chile, and the Phoenix
cryogenic echelle spectrograph.  The first and last spectra had spectral
resolving power of $\sim$70,000, while the other two had resolving power of
$\sim$50,000.  Although V2116~Oph was very challenging to find and
observe at KPNO and MSO, observations with the Gemini South telescope
were straightforward because the target was acquired with the help of a
guide star.  At $K\sim$~8 mag V2116~Oph is a relatively bright star
for an 8\,m telescope.  The spectra obtained with Phoenix on Gemini
South are of excellent quality and were useful in abundance analysis
as well as the determination of the orbit.

Standard observing and reduction techniques were used \citep{j92}.
Wavelength calibration posed a challenge, because the spectral coverage
was far too small to include a sufficient number of ThAr emission lines
for a dispersion solution.  Our approach was to utilize absorption
lines in a K~III star to obtain a dispersion solution.  Several sets of
lines were tried, including CO, Fe~I, and Ti~I.  These groups all gave
consistent results.

Radial velocities of the program star were determined with the IRAF
cross-correlation program FXCOR \citep{f93}.  The reference star was
$\delta$~Oph, an M-giant IAU velocity standard, for which we adopted a
radial velocity of $-$19.1 km~s$^{-1}$ from the work of \citet{sbf90}.

\section{ORBITAL ELEMENTS}

To determine the orbital period, we first fit a sine curve to our 17
velocities for trial periods between 100 and 1500 days with a step size
of 0.1 days.  For each period the sum of the squared residuals was
computed, and the period with the smallest value of that sum, 1154.5
days, was identified as the preliminary value of the orbital
period.  We note that a phase plot of our velocities, determined with
the previously suggested orbital period of 304 days \citep{cetal86,
petal99}, shows that both maximum and minimum velocities occur at the
same phases.  Thus, the 304-day period is clearly excluded.
\citet{cr97} speculated that the 304 day period was in fact an artifact
of the strong 1/f torque noise in the pulsar's spin behavior.

Adopting the 1154.5-day period and unit weight for all velocities,
initial orbital elements were computed with BISP, a computer program
that implements a slightly modified version of the Wilsing-Russell
method \citep{whs67}.  The orbit was then refined with SB1
\citep{bel67}, a program that uses differential corrections.  The best
fitting period is 1160.8 d or 3.18 yr.  Because of the relatively low
orbital eccentricity of 0.101 $\pm$ 0.022, we computed a circular-orbit
solution with SB1C (D. Barlow 1998, private communication), which also
uses differential corrections.  The tests of \citet{ls71} indicate that
the eccentric solution is to be preferred.  Orbital phases for the
observations and velocity residuals to this final solution are given in
Table~1.  In Figure~1 the velocities and computed velocity curve are
compared, where zero phase is a time of periastron passage.

The period of 3.18 years has resulted in large phase gaps since only
1.5 orbital periods have been covered.  Nevertheless, the elements,
listed in Table~2, are reasonably well determined.  The mass function,
which is the minimum mass of the unseen star, is quite large,
0.371$M_{\sun}$, and thus consistent with the secondary being a neutron
star.  The center-of-mass velocity of $-$177 km s$^{-1}$ is within the
range of emission line velocities of $-$120 to $-$370 km s$^{-1}$ that
\citet{cvkl98} found for this system.  The standard error of an
individual velocity is 0.85~km\,s$^{-1}$ which is in good agreement
with uncertainties derived in our work on other S-type symbiotic
systems \citep{fetal00}.  V2116 Oph is classified as a variable X-ray
source, not a late-type variable star, in the GCVS and there is no
indication of intrinsic velocity variation in the M III star.  The
orbital elements presented here supersede the preliminary results of
\citet{hetal03}.  The ephemeris for a possible eclipse of the neutron
star is \begin{displaymath} T_{conj} = HJD~2,452,235.8(\pm 53) +
1161(\pm 12)E.  \end{displaymath} 
where E is the integer number of 1161-day cycles after the given time
of conjunction.  This ephemeris predicts an upcoming mid-eclipse
on local date 2008 March 31.

\section{ABUNDANCES}\label{abunds_sect}

The high-resolution infrared spectra from which radial velocities were
determined were also used to derive abundances for a small number of
elements.  With limited spectral coverage, the following
atomic and molecular species were analyzed: Fe I, Na I, Sc I, Ti I,
$^{12}$C$^{16}$O, $^{12}$C$^{17}$O, $^{12}$C$^{14}$N, $^{16}$OH, and
H$^{19}$F, with resulting abundances for Fe, $^{12}$C, $^{14}$N,
$^{16}$O, $^{17}$O, $^{19}$F, Na, Sc, and Ti.  The line selection (with
these lines being largely unblended and measurable in terms of
equivalent widths) is listed in Table~3, along with excitation
potentials, gf-values, and equivalent widths.  Although equivalent
widths are listed in this table, all abundances were derived ultimately
from spectrum synthesis.

In a stellar abundance analysis, the model atmosphere parameters of
effective temperature (T$_{\rm eff}$), surface gravity (parameterized
by log~g, with g in units of cm\,s$^{-2}$), microturbulent velocity
($\xi$), and metallicity must be specified.  Because V2116~Oph has a
very large and uncertain reddening, photometric colors cannot
be used to derive an accurate T$_{\rm eff}$.  We rely instead on the
spectral type, as discussed by \citet{cr97}, and use this as an initial
temperature estimator.  \citet{cr97} determined a range of spectral
classes from M3 to M6, however, their various spectral indicators peak
near M4 to M5.  This suggests that the effective temperature for
V2116~Oph falls near T$_{\rm eff}$ = 3400K, with an uncertainty of
about $\pm$200K.

A series of model atmospheres, covering a range of effective
temperatures (T$_{\rm eff}$ = 3200 -- 3600K) and gravities (log~g = 0.1
-- 1.0), was generated with a version of the MARCS code
\citep{getal75}.  The $^{12}$C$^{16}$O lines span a range of excitation
potential and line strength.  Thus, they can be used to constrain both
T$_{\rm eff}$ and the microturbulent velocity ($\xi$) by requiring no
trends in the derived carbon abundance with both $\chi$ and reduced
equivalent width.  This exercise indicates that the most consistent set
of stellar parameters to describe V2116 Oph, based upon the infrared
lines, is T$_{\rm eff}$ = 3400K, log~g = 0.5, and $\xi$ = 2.4
km~s$^{-1}$.  These parameters are typical for a cool, low- to
intermediate-mass, first ascent giant or asymptotic giant branch (AGB)
star.

A near-solar metallicity is indicated, so final model atmospheres were
generated with solar metallicities.  A sample comparison of synthetic
spectra with the observed spectrum is shown in Figure~2, where the HF
1--0 R9 line, a $^{12}$C$^{17}$O line, and a $^{12}$C$^{16}$O line are
shown.  Abundances are listed in Table~4, with the uncertainties
estimated by changing stellar parameters by $\pm$100K in T$_{\rm eff}$,
$\pm$0.3 dex in log~g, and $\pm$0.3 km--s$^{-1}$ in $\xi$.  As can be
seen, the abundances of Fe, $^{16}$O, $^{19}$F, Na, Sc, and Ti are
slightly subsolar, while $^{12}$C is somewhat lower in abundance and
$^{14}$N is enhanced.

The lowered $^{12}$C and elevated $^{14}$N are the expected result
of the red-giant first dredge-up phase.  Indeed, the $^{12}$C and
$^{14}$N do not fall far from a CN-mixing line of nitrogen versus
carbon starting from initial solar-like C and N abundances.  These
results also agree with the lower limits derived for the
$^{16}$O/$^{17}$O ratios of $\le$1500, which is a value expected for a
low-mass (M$\sim$1.0--1.3M$_{\odot}$), first-ascent giant
\citep[see the review of][]{smith90}. The indication from the $^{17}$O
abundance that V2116~Oph is a low-mass giant is in agreement with the
result derived from the mass function. 

\section{DISCUSSION}

\subsection{Evolutionary State of the M Giant}

Based on X-ray and infrared observations, \citet{cr97} constrained the
basic properties of the M giant and the distance to the V2116 Oph
system.  \citet{cr97} then discussed the evolutionary state of the giant,
concluding that it is either a low-mass star near the tip of the
first-ascent giant branch or a low- or intermediate-mass star beginning
its ascent of the AGB.  Using luminosity and extinction arguments, they
suggested that the giant is near the tip of the first-ascent red-giant
branch.

Our results enable us to place additional observational constraints on
the late-type giant.  The large value of the mass function
significantly restricts the mass of the M giant because of constraints
on the neutron star mass.  The upper limit to the mass of a neutron
star is $\sim$3 M$_{\sun}$, when the internal sound speed reaches the
speed of light.  A maximum neutron star mass of 3 M$_{\sun}$ results in
a maximum mass of 5.5 M$_{\sun}$ for the M giant (Fig.~3).  However,
such a large mass for the neutron star seems unlikely.  The masses of
neutron stars in binary radio-pulsar systems are all very close to 1.35
M$_{\sun}$ \citep{tc99}, and so we adopt a mass of 1.35 M$_{\sun}$ for
the neutron-star component in V2116~Oph.  With a neutron star mass of
1.35 M$_{\sun}$, the value of the mass function requires that the mass
of the M~giant be $\leq$1.22 M$_{\sun}$, consistent with a low-mass
star.  A mass of 1.22 M$_{\sun}$ requires an edge-on orbital
inclination.  The requirement that the lower mass star have time to
evolve to a giant implies that the inclination can not be near to
face-on.  For example, as the inclination is decreased from edge-on
(90$^\circ$) a neutron star mass of 1.35 M$_{\sun}$ and an inclination
of 70$^\circ$ result in a mass of 1.0 M$_{\sun}$ for the M giant, while
an inclination of 60$^\circ$ results in an M giant mass of 0.72
M$_{\sun}$.

\citet{cr97} examined a number of optical and near-infrared spectra of
V2116~Oph.  They suggested that optical flux from the neutron star
accretion disk and from X-ray heating of the red giant weakens the TiO
band strengths much of the time, making the spectral class appear as
early as M3.  They determined a mean spectral class of M5, while a
calculation of the X-ray luminosity suggests a M6 spectral class,
corresponding to $T_{eff}$ = 3380K.  This result is in excellent
agreement with our value of 3400K, determined from spectrum synthesis
of our infrared spectra.

Using $T_{\rm eff}$ = 3400K and $R = 103$ R$_{\sun}$ (\S5.2) with the
equation $L = 4\pi\sigma R^{2} T_{\rm eff}^{4}$, we derive a luminosity
of 1270 L$_{\sun}$ for the M~giant.  This luminosity puts the red giant
very near the tip of the first-ascent giant branch, or on the lower
portion of the AGB \citep{cetal96}.  A first-ascent giant is in
agreement with the C and N elemental abundances and the oxygen isotopic
abundances.  The above temperature and luminosity combined with an
A$_v$=5 mag (\S1) results in a distance of 4.3 kpc.  In spite of its
position in the sky, which is quite near the direction to the Galactic
center, V2116 Oph is clearly not an object associated with the center
of the Milky Way.

The models of \citet{cetal96} indicate that that the current mass of
$\leq$1.22 M$_{\sun}$ corresponds to a main sequence mass of $\leq$1.34
M$_{\sun}$ and that such a star with solar metallicity takes $\sim$5 $\times$
10$^9$ yrs to reach the first ascent giant branch tip.  Assuming the
neutron star resulted from a core collapse supernova with a M $\geq$ 10
M$_{\sun}$ progenitor, the progenitor lifetime up to core collapse was
a few times 10$^7$ years.  If true, the neutron star would be 5 billion
years old, a point we will return to below.

\subsection{Synchronization and Circularization}

\citet{z77} has shown that tidal forces in binaries cause rotational
synchronization and orbital circularization, and that the time scale
for synchronization is shorter than that for circularization.  Both the
synchronization and circularization times are principally dependent on
the ratio of the semimajor axis of the relative orbit, $a$, to the
giant star radius, $R$.  The circularization time scales as $(a/R)^8$
while the synchronization time scales as $(a/R)^6$ \citep{sok00}.  The
orbit of the red giant in the V2116~Oph system provides the value for
the semimajor axis of the red giant times the sine of the inclination,
a$_g~{\rm sin}~{\it i}$.  Designating the semimajor axis of the neutron star's
orbit as a$_n$, the semimajor axis of the relative orbit is
$a\,=\,a_g~+~a_n$.  The masses of the components are approximately equal
so for this system we assume $a\,=\,2a_g$.

Based on the work of \citet{z77}, \citet{setal94} and \citet{metal00}
have argued that in most S-type symbiotics, the giant is synchronously
rotating.  Indeed, with the orbital parameters and masses derived for
V2116~Oph, the tidal synchronization time (the time required to spin up
the red giant to synchronous rotation with the orbit) is only $\sim$
10$^4{\sin}^{-6}~{\it i}$ years, according to the formulae in \citet{sok00}.
The orbit of V2116~Oph has a modest eccentricity.  Thus, the rotational
angular velocity of the M giant will synchronize with that of the
orbital motion at periastron to achieve pseudosynchronous rotation
\citep{h81}.  With equation (42) of \citet{h81} we calculated a
pseudosynchronous period of 1075.5 days.  Although the synchronization
and circularization calculations involve raising the ratio $(a/R)$ to
large powers, it is very unlikely that these times can be a factor of
10 larger.  The semimajor axis is known to 2\%.  As discussed below,
our value for the radius of the giant, 103 R$_\odot$, is uncertain by
$\sim$30\%.

If, as suggested by theory, the M giant is pseudosynchronously rotating,
its projected rotational velocity can be used to estimate its minimum
radius.  To determine {\it v}~sin~{\it i} of the M giant, we measured
the full-width at half-maximum of a few atomic features at 2.2~$\mu$m.
We also measured the same lines in several late-type giants with known
{\it v}~sin~{\it i} values.  With the latter set of stars, we produced
an empirical broadening calibration similar to that of \citet{f97}.
Using the calibration and an adopted macroturbulence of 3~km~s$^{-1}$, 
we determined {\it v}~sin~{\it i} = 8 $\pm$~1~km~s$^{-1}$ for
the M giant.  The pseudosynchronous period and the {\it v}~sin~{\it i}
value result in a {\it minimum} radius (i.e., sin~{\it i} = 1) 
of 170 $\pm$ 21 R$_{\sun}$.

Another way of estimating the red-giant radius is to use the results of
\citet{ds98}.  They computed radii for a sample of nearby M giants that
have relatively well determined Hipparcos parallaxes.  They found a
median radius of $\sim$150~R$_{\sun}$ for the M6 giants in their sample
but their Figure~6 shows that the radius of a given spectral type is
mass dependent.  Our orbital and abundance results indicate that the M
giant has a mass $\lesssim$ 1.2 M$_{\sun}$.  Figure~6 of \citet{ds98}
shows that the M6 giants with the lowest masses, 1.0 M$_{\sun}$ $\leq$
M $\leq$ 1.6 M$_{\sun}$, have a mean radius of 118 R$_{\sun}$,
significantly smaller than the minimum radius for pseudosynchronous
rotation.

The best determination of the M~giant radius is computed from our
spectroscopic determination of log~g = 0.5 $\pm$ 0.3 and a maximum mass
for the red giant of 1.22 M$_{\sun}$, which result in a radius of 103
R$_{\sun}$.  The estimated uncertainty of log~g results in a radius
range of 73 to 145 R$_{\sun}$.  Decreasing the mass of the red giant,
decreases its radius.  So, the mean radius from the low-mass M6 giants
of \citet{ds98} is consistent with our spectroscopic determination.
Contrary to theoretical expectations, both those values argue that the
M~giant is not rotating pseudosynchronously.  \citet{petal97} find a
value for the angle between the observer line of sight and the neutron
star spin axis of 56\arcdeg $\pm$ 8\arcdeg.  Since the M giant is not
rotating pseudosynchronously, we are unable to present arguments that
the spin axis of the neutron star should be normal to the orbital
plane.

The V2116~Oph orbit has significant eccentricity.  The time scale for
circularization of the orbit caused by tides induced in the red giant
by the companion is $\sim 5\times10^6\sin^{-8}~{\it i}$ yr from the relations
given by \citet{sok00}.  Since mass loss and transfer as well as tides
tend to circularize orbits \citep{htp02}, a kick from the supernova
explosion which formed the neutron star is one way to produce a
non-circular orbit.  The apparent lack of ejecta seems to exclude a
relatively recent supernova explosion, although the reddening toward
V2116~Oph is large and it is possible that supernova ejecta surrounding
the system have not been detected.  However, there are many red giants
currently in very close binary systems that appear to be in eccentric
orbits \citep{sos04}.  Thus the eccentricity of the V2116 Oph orbit
does not necessarily limit the time since the supernova event. \citet{sok00}
has argued that the eccentricity in such evolved systems is caused by
an enhanced mass-loss rate during periastron passage.  Indeed, other
articles of this series \citep{fetal00,f00b,f01} have found symbiotic
systems with eccentric orbits for periods as short as 450 days;
however, these systems contain a white dwarf rather than a neutron star
companion to the M giant and are less massive.

The X-ray flux and its behavior require the presence of an accretion disk
around the neutron star.  \citet{cr97} favored a stellar wind rather
than Roche-lobe overflow to form the accretion disk.  Placing V2116~Oph
on the first-ascent giant branch implies that the M giant does not fill
its Roche lobe.  This is in agreement with less direct but powerful
arguments presented by \citet{cr97}, based on the stellar radius and
luminosity as determined from the V2116~Oph reddening, variability, and
stellar wind and the GX1+4 X-ray luminosity and accretion torque.  The
optical variability of V2116~Oph and the cessation of flickering from
the presumed accretion disk during faint intervals suggest that the
mass transfer from the red giant to the X-ray source is time variable
\citep{jetal97}, as would be the case for a non-Roche lobe filling
system.  Although the orbit of V2116~Oph is not circular, the
eccentricity is low, and we can determine an approximate Roche-lobe
radius.  With Kepler's third law and the formula of \citet{egg83}, we
find a Roche-lobe radius = 236 R$_{\sun}$.  Comparison with our derived
giant radius of 103 R$_{\sun}$ indicates that like most S-type
symbiotic binaries \citep{ms99}, the M giant is far from filling its
Roche lobe.  Thus, the mass transfer to the neutron star does 
indeed result from the M giant stellar wind
rather than Roche-lobe overflow.

\subsection{Evolution of the Binary System}

The existence of X-ray binaries is proof of what might seem an unlikely
proposition, that a binary system can survive a supernova explosion.
In the case of V2116~Oph the current M giant not only survived the
explosion but did so with no obvious change in the surface abundances.
Surprisingly, there is no reason to expect 
the abundances of what is now the
M~giant to be altered significantly by the supernova explosion.  If 
the stars had been in
contact, the area subtended by the then main sequence star was on the
order of 0.001 percent of the surface area of the supernova precursor.
Even if the supernova ejecta exceeded several solar masses of oxygen,
the amount accreted by the main sequence star would have resulted in
undetectable surface abundance changes after the main sequence surface 
has been subducted in the giant stage.

Defining characteristics of high-mass and low-mass X-ray binaries are
given by \citet{vh95}.  In low-mass X-ray binaries, the companion to
the compact object has a mass $\lesssim$ 1.0 M$_{\sun}$.  These systems
belong to a very old stellar population and do not show runaway
characteristics.  Based solely on our upper limit to the mass of the
red giant, V2116~Oph/GX~1+4 would marginally meet the criteria for a
low-mass system.  However, the abundances of the red giant are nearly
solar and hence the system does not belong to a very old population.
The center-of-mass velocity of the system is quite large, $-$177
km~s$^{-1}$.  While the V2116~Oph system fits the mass criteria for
low-mass systems, the near-solar abundances make V2116~Oph a deviant
member.

Furthermore, the large velocity of V2116~Oph suggests that it is a
runaway system.  On 2003 February 16 the 1.564 $\mu$m spectrum of a
star separated from V2116~Oph by 19 arcseconds, Star 6 in \citet{cr97}
Table~2, was observed with a spectral resolving power of 50000 with
Phoenix at Gemini South.  This star, which has a similar visual
magnitude to V2116~Oph, proved to be an M star with a velocity of
$+$118 km~s$^{-1}$.  The observation of one star of unknown distance
does not demonstrate the kinematics of the V2116~Oph field, but the 295
km~s$^{-1}$ velocity difference between the two systems suggests that a
significant velocity was imparted to the V2116~Oph system by the
supernova explosion, giving it a runaway velocity.

Considerable discussion exists in the literature on ways to produce
X-ray binaries.  As reviewed by \citet{cil90}, \citet{ity95},
\citet{vh95}, and others, while there is consensus that high mass
systems are the result of massive star evolution, four schemes have
been advanced for the formation of low-mass X-ray binary systems.
These four are:  (1) capture of a neutron star by another star in a
dense stellar system (e.g. globular cluster), (2) the direct evolution
of one star of a primordial binary into a neutron star, (3) triple star
evolution of a system including a massive star binary, and (4) a low
mass binary where one member first becomes a white dwarf then undergoes
mass accretion resulting in collapse to a neutron star.

For a field star, tidal capture by a neutron star is unlikely.  The
neutron star must approach the star to be captured by 2 to 3 times the
target star radius, although the odds may be increased by involving a
binary system \citep{cil90}.  Since the neutron star and the M giant
have nearly equal masses the capture conditions are very limited.  The
tidal capture mechanism is most likely for stars in the cores of
globular clusters.  The abundances of V2116~Oph do not rule out a
scenario where V2116~Oph was formerly a member of a low mass globular
cluster and escaped as the result of being captured by a runaway
neutron star.  Achieving escape velocity from a globular cluster is a
major difficulty, but this scenario is remotely possible \citep{k84}.
Nonetheless we discount tidal capture for the formation of the
V2116~Oph system.

Evolution of V2116~Oph from a massive star -- low mass star binary is
tightly constrained by the requirement that a bound binary system
survive a supernova event.  Standard models show that systems ejecting
half or more of their total mass in a supernova explosion become
unbound \citep{dc87}.  The production of a neutron star remnant from a
main sequence star requires a $\sim$ 9 M$_{\sun}$ main sequence star.
Without any velocity kick, keeping the system bound during a supernova
with a massive star and a 1.3 M$_{\sun}$ companion requires that the
neutron star progenitor have a mass no greater than $\sim$4 M$_{\sun}$
at the time of the supernova event.  Certainly considerable mass loss
can occur before the supernova progenitor explodes.  Calculations by
\citet{mm88} for single star mass loss indicate that a 9 M$_{\sun}$
star could be reduced to 3.5 M$_{\sun}$ before the star explodes.
Furthermore, \citet{h83} suggests that in unusual conditions of orbital
phase and eccentricity a system loosing more than half of the total
mass can survive.  The presence of a velocity kick from the supernova
further constrains the mass loss to keep the system bound and this
scenario for V2116 Oph seems unlikely.

A modification to the direct evolution scheme invokes common envelope
evolution \citep{ts00} which would allow a binary with high mass and
low mass members to survive the supernova phase.  A common envelope
stage when the massive star becomes a supergiant can lead to the
ejection of the common envelope, resulting in a binary composed of an
evolved stellar core and a main sequence companion.  The reduced mass
difference between the evolved core and the main-sequence star allows
the system to remain bound.  The orbit following the supernova
explosion would be highly eccentric \citep{dc87} but tidal interaction
would rapidly circularize the orbit.

\citet{cil90} and \citet{it99} note that it is possible to form a
V2116~Oph like binary through common envelope evolution and merger
starting with a triple system.  One such scenario invokes a massive
close binary with a distant dwarf companion.  The massive binary
undergoes mass transfer and a supernova.  The remaining massive
star--neutron star system then undergoes common envelope evolution and
a merger forming a Thorne-Zytkov object, i.e. a red giant with a
neutron star core.  This is followed by a common envelope phase with
the dwarf during which the neutron star envelope is ejected leaving a
neutron star-dwarf system.  Starting from either a binary or a triple
system and by invoking various special conditions it does appear
possible to evolve a high mass -- low mass binary into a neutron star
-- low mass binary.

The last scenario is accretion-induced collapse (AIC) of a white
dwarf.  The more massive member of a low mass binary system initially
becomes a white dwarf.  Mass transfer then takes place from the initial
secondary, resulting in collapse of the white dwarf to a neutron star.
The appeal of the model is that small mass ejection is involved and
hence minimal disruption to the orbit.  An initially massive white
dwarf and a high rate of mass accretion are required.  Broad objections
to AIC concern the details of the mass transfer and the requirement
that the supernova event leave a remnant.  The
accretion-induced collapse scenario is reviewed at length by
\citet{cil90}.  A consequence of the AIC is that the stars in the
system can have similar ages.

Objections are possible when applying any of these scenarios to the
V2116~Oph system.  However, in many aspects V2116~Oph appears to be a
perfect example of a AIC system as suggested by \citet{vdh84}.  The
secondary is only now on the giant branch.  If the dwarf in the prior
dwarf-white dwarf star system had nearly filled its Roche lobe, a
high-rate mass transfer episode would have taken place as post-main
sequence evolution commenced.  The accretion rate from a 1.2 M$_\sun$
secondary is optimal for AIC \citep{vdht84}.  \citet{vdht84} also point
out that a precursor white dwarf $\sim$10$^9$ years old would have the
optimal interior structure for AIC.  A binary of the age of V2116~Oph
could have a white dwarf this old.  Furthermore, the current
evolutionary state of V2116~Oph, i.e.~a first ascent red giant and a
neutron star companion would be the product of a relatively recent AIC
event.  The evolution of the binary orbit following the supernova
results from the explosive ejection of $\sim$0.2 M$_\sun$.  This
enlarges the binary orbit but does not disrupt the system.  The ejecta
from such a supernova would likely be mainly helium \citep{nomoto82}
and would not impact the measured abundances of the current M III.

The massive star and AIC scenarios produce neutron stars with ages
differing by billions of years.  If it were possible to date the
neutron star it would be possible to exclude certain classes of
models.  The neutron star spin rate does not provide any insight since,
as noted in the introduction, the neutron star rotation rate of
V2116~Oph is linked to the mass transfer.  The surface temperature of
the neutron star reflects the interior structure and hence age
\citep{bp79}.  This measurement would be difficult to carry out due to
mass transfer in the V2116~Oph system.  However, \citet{tvdh86}
proposed that the surface magnetic fields of neutron stars decay on a
time scale measured in 10$^6$ years.  The neutron star in V2116~Oph has
a magnetic field that is among the strongest known, implying youth.  At
odds with \citet{tvdh86}, \citet{pv91} noted evidence that neutron star
fields do not undergo massive decay, but did find that field decay of
at least a factor of four over a period of $\sim$10$^{10}$ years is
required to match neutron star field statistics.  The field of
V2116~Oph is near the maximum limit supportable in a neutron star \citep{fr77}
and can not possibly be the age of the M giant component in the
system.

Given the nature of the V2116~Oph system, with the extraordinary
presence of a neutron star in a 3-year orbit, near solar abundances of
the M-giant, and the large radial velocity, the binary is the only
system of its type known in the Milky Way galaxy. This suggests either
a very rapid evolutionary state or a highly unusual formation
mechanism.  While the giant phase is not short lived, it is possible
that the system is currently undergoing a short lived phase of high
mass transfer associated with the late-type star's evolution to the red
giant tip.  The probabilities associated with the formation mechanism
are best summarized by \citet{ity95} who notes that a binary evolving
to this end must escape both the hazards of merger and disruption from
supernova explosion.

The future evolution of V2116~Oph is an interesting topic for speculation.
When the current red giant becomes an AGB star in $\sim$10$^8$ years,
single star evolution predicts that it will be much larger in size
than its current orbit \citep{cetal96}.  Furthermore, tidal interaction
and perhaps magnetic braking will reduce the current separation.
Thus, expansion of the giant will result in the V2116~Oph system becoming
a contact binary.  An AGB star does not contract as a
result of mass loss, so as the AGB star continues to expand the neutron
star will be engulfed in the envelope of the AGB star and co-rotation,
the requirement for escaping common envelope evolution, will be lost.
Common envelope systems rapidly eject mass and the evolutionary end
product will be a neutron star -- white dwarf binary similar to PSR
0655+64 or PSR 0820+02 \citep{vdht84}.  The existence of these objects 
suggests, as do arguments about the lifetime of the mass transfer stage,
that the evolutionary path of V2116 Oph is not extremely rare
in the Milky Way.  V2116 Oph is notable because the current mass transfer
rate makes it highly visible.

\section{CONCLUSIONS} 
 
The determination of a time series of infrared radial velocities of the
M giant V2116~Oph have confirmed the association of this star with the
neutron star/X-ray source GX 1+4.  The single-line spectroscopic binary
orbit is now well determined and at 1161 days exceeds the orbital
period of any other low-mass X-ray binary by a factor of nearly 50
\citep{lph01}.  While the inclination is unknown, based on other X-ray
binary systems the mass of the neutron star is probably $\sim$ 1.35
M$_{\sun}$.  The plane of the orbit must lie close to the line of
sight, implying that eclipses of at least part of the mass transfer
flow or accretion disk are possible.  The next predicted date of
``mid-eclipse'' is 2008 Mar 31.  The abundances of the M giant, as well
as the mass function of the orbit combined with the adopted mass of the
neutron star, indicate a mass of $\sim$ 1.2 M$_{\sun}$ for the M
giant.  The abundances of the M giant also suggest that it is a first
ascent giant, in agreement with arguments presented in the detailed
study of the system by \citet{cr97}.  Our radius of the M giant and the
X-ray behavior of the system indicate that the giant does not fill its
Roche lobe, a conclusion consistent with its identification as a first
ascent giant.  The fact that the giant does not fill its Roche lobe,
makes V2116~Oph very different from other low-mass X-ray binaries.

The evolutionary history of this and other neutron star binaries is an
area of continued speculation.  Should the neutron star have originated
with the supernova explosion of a $\sim$10 M$_{\sun}$ star, the neutron
star is ancient, having co-existed with the progenitor of the current M
giant for nearly 5 billion years.  However, the activity of the
V2116~Oph neutron star strongly suggests that it was formed 10$^7$
years ago or less.  Since we know the age of the M giant is $\sim$5 x
10$^9$ years, this implies that a massive star never existed in this
system.  This strongly suggests that the neutron star resulted from
accretion induced collapse.  The pre-collapse system consisted of a
white dwarf and the first ascent giant nearly filling its Roche lobe.

If the accretion induced collapse scenario is correct, the V2116~Oph
neutron star is a fairly recent remnant from a single degenerate type I
supernova event.  To an observer prior to the V2116~Oph supernova, the
white dwarf -- red giant system would have appeared as a symbiotic
system.  There are few symbiotic systems known that contain high mass
white dwarfs and none that also contain a companion close to filling
its Roche lobe.  However, the fraction of symbiotic binaries that have
been fully characterized is small.  Continued exploration of these
systems may reveal a pre-accretion induced collapse system.

\acknowledgements

This paper is based in part on observations obtained at the Gemini
Observatory, which is operated by the Association of Universities for
Research in Astronomy, Inc., under a cooperative agreement with the NSF
on behalf of the Gemini partnership: the National Science Foundation
(United States), the Particle Physics and Astronomy Research Council
(United Kingdom), the National Research Council (Canada), CONICYT
(Chile), the Australian Research Council (Australia), CNPq (Brazil),
and CONICRT (Argentina).  The observations were obtained with the
Phoenix infrared spectrograph, which was developed and is operated by
the National Optical Astronomy Observatory.  The spectra were obtained
as part of programs GS-2003A-DD-1 and GS-2004A-DD-1. 
 
The anonymous referee provided useful insight into the synchronization
and circularization time scales.
We thank R. Blum for carrying out some of the Gemini South
observations.  This research has been supported in part by NASA grants
NCC5-511 and NSF grant HRD-9706268 to Tennessee State University.  TL
was funded by the Austrian Academy of Sciences (APART) and the Austrian
Science Fund (FWF-project P18171).  We have made use of the SIMBAD
database, operated by CDS in Strasbourg, France, as well as NASA's
Astrophysics Data System Abstract Service.

\clearpage

\singlespace 
\begin{deluxetable}{ccccrc} 
\tablenum{1} 
\tablewidth{0pt} 
\tablecaption{RADIAL VELOCITIES OF V2116 OPH} 
\tablehead{\colhead{Hel. Julian Date} & \colhead{Wavelength} & \colhead {} &  
\colhead{Velocity} & \colhead{$O-C$} & \colhead{} \\ 
\colhead{(HJD $-$ 2,400,000)} & \colhead{($\mu$m)} & \colhead {Phase} &  
\colhead{(km~s$^{-1}$)} & \colhead{(km~s$^{-1}$)} & \colhead{Source} 
} 
\startdata 
 51,362.778 & 1.5612--1.5682 & 0.500  & $-$162.5  &     1.4  & KPNO 4\,m \\ 
 51,363.736 & 1.5612--1.5682 & 0.501  & $-$163.9  &  $-$0.0  & KPNO 4\,m \\ 
 51,677.953 & 1.5534--1.5606 & 0.771  & $-$174.3  &  $-$0.2  & KPNO 4\,m \\ 
 51,737.750 & 1.5590--1.5662 & 0.823  & $-$178.3  &     0.6  & KPNO 2.1\,m\\ 
 51,982.019 & 1.6536--1.6610 & 0.033  & $-$194.2  &  $-$1.4  & KPNO 4\,m \\ 
 52,048.137 & 1.6190--1.6250 & 0.090  & $-$190.5  &     0.7  &   MSO      \\ 
 52,098.932 & 1.6275--1.6330 & 0.134  & $-$188.0  &     0.4  &   MSO      \\ 
 52,133.982 & 1.6275--1.6330 & 0.164  & $-$186.3  &  $-$0.3  &   MSO      \\ 
 52,155.957 & 1.6275--1.6330 & 0.183  & $-$184.3  &     0.1  &   MSO      \\ 
 52,357.272 & 1.6275--1.6330 & 0.356  & $-$170.1  &  $-$0.2  &   MSO      \\ 
 52,399.268 & 1.6275--1.6330 & 0.393  & $-$168.2  &  $-$0.5  &   MSO      \\ 
 52,447.177 & 1.6275--1.6330 & 0.434  & $-$165.5  &     0.2  &   MSO      \\ 
 52,502.980 & 1.6275--1.6330 & 0.482  & $-$165.0  &  $-$0.8  &   MSO      \\ 
 52,749.900 & 2.2212--2.2315 & 0.695  & $-$167.8  &     0.5  &   Gemini S \\ 
 52,771.672 & 2.3304--2.3413 & 0.714  & $-$170.7  &  $-$1.2  &   Gemini S \\ 
 52,849.503 & 2.2900--2.2980 & 0.781  & $-$174.9  &     0.0  &   Gemini S \\ 
 53,098.811 & 2.2212--2.2315 & 0.995  & $-$191.8  &     0.5  &   Gemini S \\ 
\enddata 
\end{deluxetable} 
 
%\clearpage 
 
\begin{deluxetable}{lc} 
\tablenum{2} 
\tablewidth{0pt} 
\tablecaption{ORBITAL ELEMENTS OF V2116 OPH} 
\tablehead{ \colhead{Parameter} & \colhead {Value }  \\ 
} 
\startdata 
Orbital Period, $P$ (days)                           & 1160.8 $\pm$ 12.4     \\ 
Orbital Period, $P$ (years)                          & 3.18 $\pm$ 0.03       \\
Time of periastron passage, $T$ (HJD)                & 2,451,943 $\pm$ 53    \\ 
Systemic velocity, $\gamma$ (km~s$^{-1}$)              & $-$176.73 $\pm$ 0.22  \\ 
Orbital velocity semi-amplitude, $K_g$ (km~s$^{-1}$)      & 14.62 $\pm$ 0.34      \\ 
Eccentricity, $e$                                    & 0.101 $\pm$ 0.022     \\ 
Longitude of periastron, $\omega_g$ (deg)            & 168 $\pm$ 17          \\ 
Projected semimajor axis, $a_g$~sin~$i$ (km)         & 232.2 $\pm$ 6.0 x 10$^6$ \\ 
Projected semimajor axis, $a_g$~sin~$i$ (R$_{\sun}$) & 334 $\pm$ 9           \\
Mass function, $f(m)$ (M$_{\sun}$)                   & 0.371 $\pm$ 0.026     \\ 
Standard error of an observation                     &                       \\ 
of unit weight (km~s$^{-1}$)                         & 0.85                  \\
\enddata 
\tablecomments{When applicable only to the red giant component of the binary 
orbit the parameters are subscripted ``g''. 
The six orbital elements have the standard definitions of
\citet{sterne41}. Definitions for $a$~sin~$i$ and $f(m)$ are given by 
\citet{batten89}.}
\end{deluxetable} 
 
\clearpage 
 
\begin{deluxetable}{lcccc} 
\tablenum{3} 
%\tablewidth{260pt} 
\tablewidth{0pt} 
\tablecaption{ABUNDANCE SPECTRAL-LINE DATA} 
\tablehead{\colhead {Species} & \colhead{Wavelength} & \colhead{$\chi$} & 
\colhead{log gf} & \colhead{EW} \\  
\colhead{} & \colhead{(\AA)} & \colhead{(eV)} &  
\colhead{} & \colhead{(m\AA) } 
} 
\startdata 
Fe I & 22,257.098 & 5.06 & $-$0.770 & 296 \\ 
     & 22,260.186 & 5.09 & $-$1.000 & 271 \\ 
Na I &  23,378.945 & 3.75 & +0.731 & 594 \\ 
Sc I & 23,266.729 & 1.43 & $-$1.280 & 539 \\ 
     & 23,404.756 & 1.44 & $-$1.278 & 581 \\ 
Ti I & 22,232.957 & 1.74 & $-$1.621 & 617 \\ 
     & 22,309.961 & 1.73 & $-$2.086 & 534 \\ 
 $^{12}$C$^{14}$N & 15,563.376 & 1.15 & $-$1.141 & 119 \\ 
 $^{12}$C$^{16}$O & 23,351.438 & 0.42 & $-$5.104 & 897 \\ 
   & 23,384.467 & 0.38 & $-$5.165 & 859 \\ 
   & 23,389.146 & 2.20 & $-$4.790 & 163 \\ 
   & 23,396.305 & 0.37 & $-$5.187 & 833 \\ 
   & 23,398.275 & 1.73 & $-$4.458 & 444 \\ 
   & 23,406.389 & 0.00 & $-$6.583 & 682 \\ 
 $^{16}$OH &  15,560.271 & 0.30 & $-$5.307 & 609 \\ 
   & 15,568.807 & 0.30 & $-$5.270 & 574 \\ 
   & 15,572.111 & 0.30 & $-$5.183 & 555 \\ 
 H$^{19}$F & 23,358.412 & 0.48 & $-$3.955 & 463 \\ 
\enddata 
\tablecomments{Molecular dissociation constants are taken to be 
D$_{\rm o}$(CN)=7.65 eV, D$_{\rm o}$(CO)=11.09 eV, D$_{\rm o}$(OH)=4.39 
eV, and D$_{\rm o}$(HF)= 5.82 eV. 
} 
\end{deluxetable} 
 
\clearpage 
\begin{deluxetable}{cccr} 
\tablenum{4} 
%\tablewidth{220pt} 
\tablewidth{0pt} 
\tablecaption{ABUNDANCES} 
 
\tablehead{ Species  & 
\multicolumn{1}{c} { V2116 Oph } & 
\multicolumn{1}{c} { Sun } & 
\multicolumn{1}{c} { [x/H] }  
} 
\startdata 
 Fe                 & 7.45$\pm$0.26 & 7.50 & $-$0.05 \\ 
 $^{12}$C           & 8.03$\pm$0.18 & 8.45 & $-$0.42 \\ 
 $^{14}$N           & 8.97$\pm$0.25 & 7.80 & +1.17 \\ 
 $^{16}$O           & 8.47$\pm$0.29 & 8.77 & $-$0.30 \\ 
 $^{16}$O/$^{17}$O  & $\ge$1500     & 2660 &  .... \\ 
 $^{19}$F           & 4.55$\pm$0.22 & 4.55 &  0.00 \\ 
 Na                 & 6.00$\pm$0.18 & 6.33 & $-$0.33 \\ 
 Sc                 & 3.18$\pm$0.15 & 3.17 & +0.01 \\ 
 Ti                 & 4.58$\pm$0.22 & 5.02 & $-$0.44 \\ 
\enddata 
\tablecomments{ Format for V2116 Oph and Solar abundances are A(x)= 
log[N(x)/N(H)] + 12.0, and [x/H]= A(x)$_{\rm Star}$ $-$ A(x)$_{\rm Sun}$. 
 } 
\end{deluxetable} 
 
\clearpage 
 
\begin{figure}[t!] 
\epsscale{0.8} 
\plotone{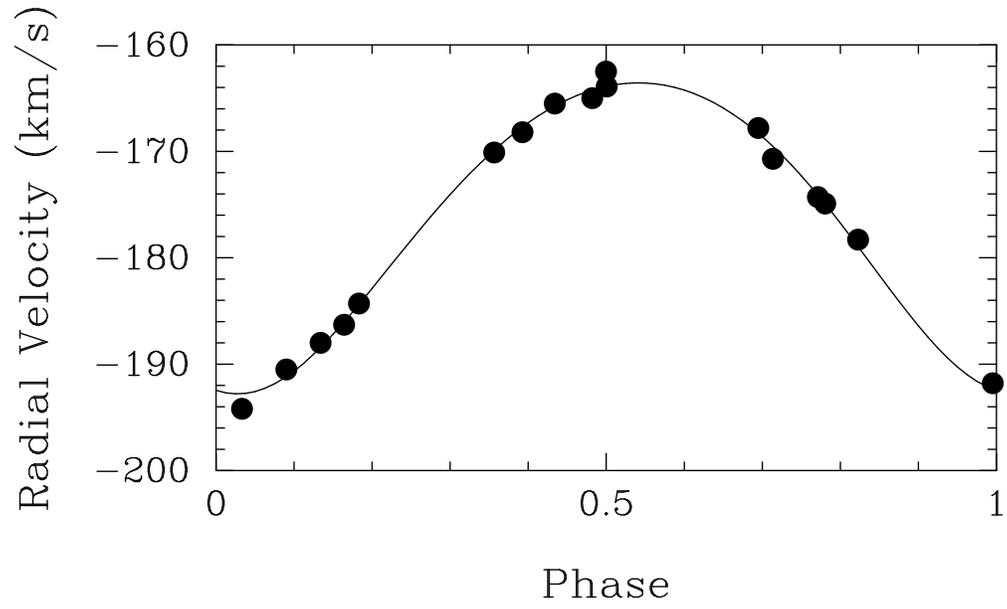} 
\figcaption{The computed radial-velocity curve of V2116~Oph compared with 
our infrared velocities.  Zero phase is a time of periastron passage.} 
\label{f:orbit}
\end{figure} 

\begin{figure}[t!] 
\epsscale{0.8} 
\plotone{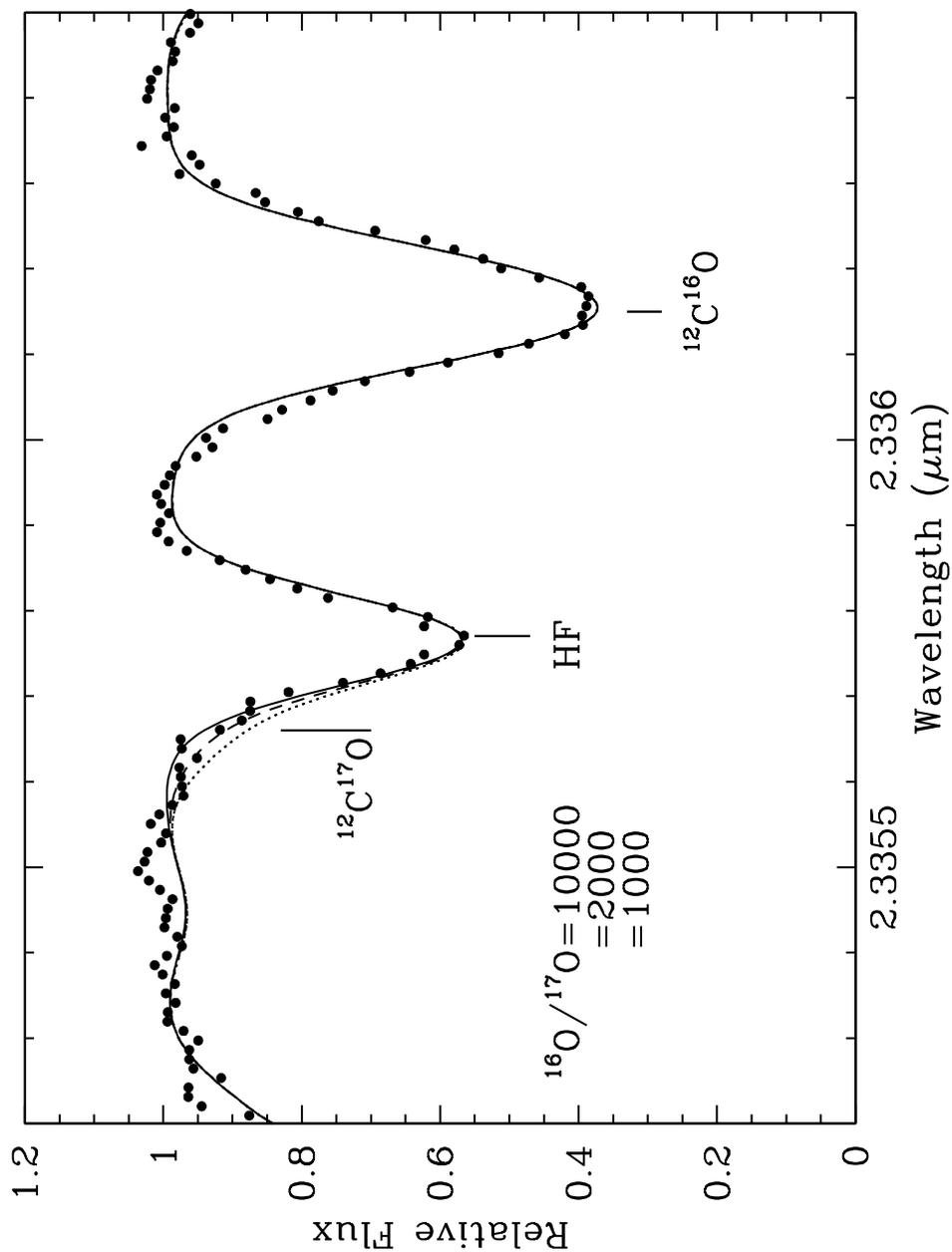} 
\figcaption{ 
Sample illustration of synthetic spectra compared to the observed
spectrum of V2116~Oph.  This region shows the 1-0 R 9 HF line, along
with a strong $^{12}$C$^{16}$O blended feature (3-1 R 24 + R 76) as
well as the expected location of a clean $^{12}$C$^{17}$O line (2-0 R
26) that is not detected.  If V2116~Oph were a first-ascent red giant
more massive than about 1.3M$_{\odot}$, the $^{12}$C$^{17}$O should be
visible.} 
\label{f:spectrum}
\end{figure}

\begin{figure}[t!]
\epsscale{0.8}
\plotone{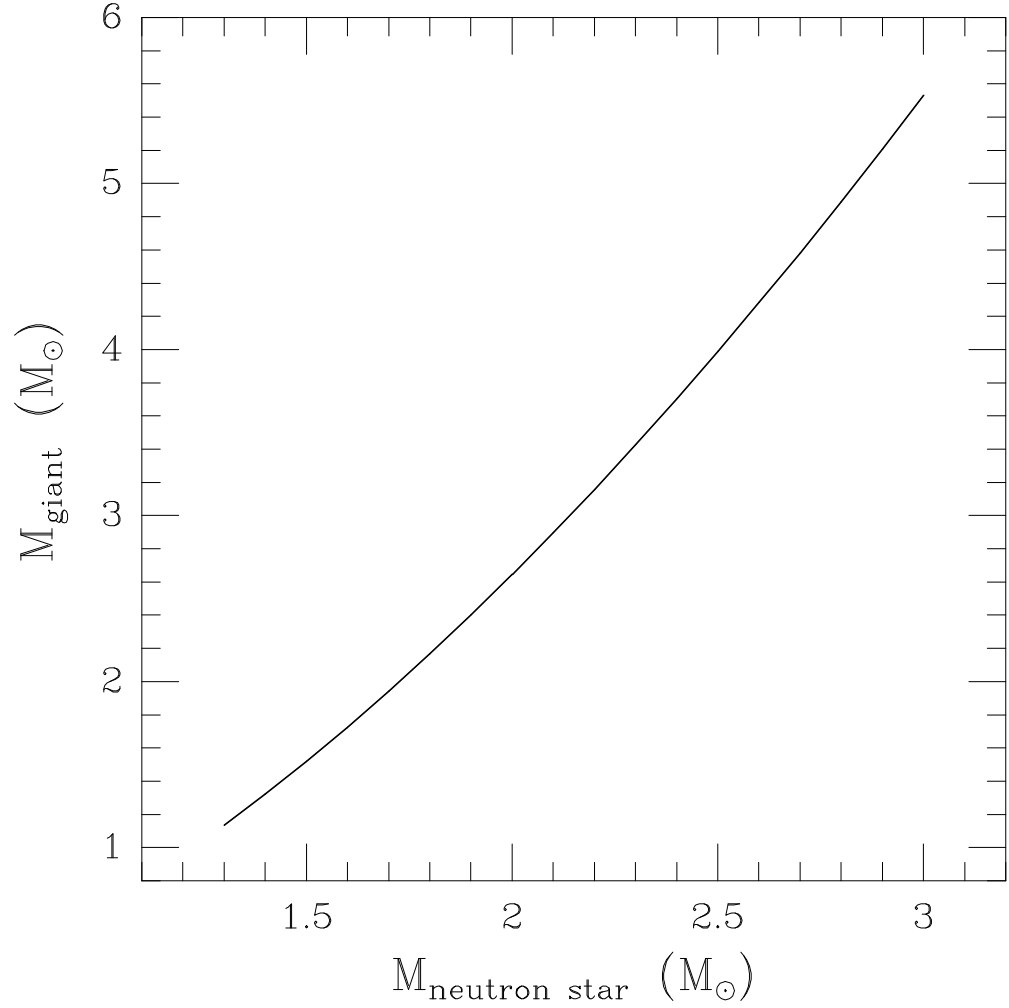}
\figcaption{
The line shows the relation between the masses, in solar units, of the
M giant and the neutron star for an orbit seen edge on.  The neutron
star mass ranges from the Chandrasekhar limit to black hole collapse.
For a given neutron star mass, decreasing the orbital inclination
decreases the mass of the M giant.  Additional constraints (see text)
suggest a neutron star mass near 1.35 M$_{\sun}$ and a nearly edge on
orbital inclination.
}  
\label{f:mass_fcn}
\end{figure}

\end{document}